\documentclass[prb,showpacs,amssymb,twocolumn,floatfix]{revtex4}
\usepackage{graphicx}
\usepackage{amsmath}
\usepackage{braket}
\usepackage{bm}

\begin{document}

\title{Wave Packet Dynamics, Ergodicity, and Localization in Quasiperiodic Chains}

\author{Stefanie Thiem}
\author{Michael Schreiber}
\affiliation{Institut f\"ur Physik, Technische Universit\"at Chemnitz, D-09107 Chemnitz, Germany}
\author{Uwe Grimm}
\affiliation{Department of Mathematics and Statistics, The Open University, Milton Keynes MK7 6AA, United Kingdom}
\date{\today}

\begin{abstract}
In this paper, we report results for the wave packet dynamics in a class of quasiperiodic chains consisting of two types of weakly coupled
clusters. The dynamics are studied by means of the return probability and the mean square displacement. The wave packets show anomalous
diffusion in a stepwise process of fast expansion followed by time intervals of confined wave packet width. Applying perturbation theory,
where the coupling parameter $v$ is treated as perturbation, the properties of the eigenstates of the system are investigated and related
to the structure of the chains. The results show the appearance of non-localized states only in sufficiently high orders of the
perturbation expansions. Further, we compare these results to the exact solutions obtained by numerical diagonalization. This shows that
eigenstates spread across the entire chain for $v > 0$, while in the limit $v \to 0$ ergodicity is broken and eigenstates only spread
across clusters of the same type, in contradistinction to trivial localization for $v = 0$. Caused by this ergodicity breaking, the wave packet
dynamics change significantly in the presence of an impurity offering the possibility to control its long-term dynamics.
\end{abstract}


\maketitle

\section{Introduction}\label{sec:introduction}

Understanding the relations between the spectral properties of a given Hamiltonian and the dynamics of wave packets which are governed by
it, remains one of the elementary questions of quantum mechanics that still poses significant challenges, further emphasized by the
discovery of quasicrystals.\cite{PhysRevLett.1984.Shechtman, PhysRevLett.1985.Ishimasa, PhysRevLett.1987.Wang} While spectra of many
Hamiltonians decompose into a point-like part and an absolutely continuous part accompanied by bounded (localized) and unbounded (delocalized or extended) eigenstates, there exists a large variety of Hamiltonians whose spectra, for certain values of the parameters, are neither pure point-like nor absolutely continuous, nor a combination of both. In this case, the spectrum contains a singular continuous part, and eigenstates are often found to be multifractal. Examples are Harper's model of an electron in the magnetic field,\cite{PhysRevB.1976.Hofstadter} the kicked
rotator\cite{PhysRep.1990.Izrailev, PhysRevLett.1992.Artuso} as well as the Anderson model of an electron in a disordered
medium.\cite{PhysRevLett.1991.Schreiber} In the case of an electron in a one-dimensional quasiperiodic system, as studied in this paper, many examples lead to spectra which are purely singular continuous as well.\cite{PhysRevB.1987.Kohmoto, JStatPhys.1989.Suto, CMathPhys.1989.Bellisard}

To address the above mentioned challenge, we investigate wave packet dynamics in one-dimensional quasiperiodic chains by numerical
simulation as well as perturbation theory, and relate the results to the hierarchical properties of these chains. The content of this paper
is organized as follows: At first we introduce the construction rule and structure of the chains in Sec.~\ref{sec:construction}. Section
\ref{sec:transport} then focuses on numerical results for the time evolution of wave packets. To obtain a better understanding of the
properties of wave packet spreading and localization, we apply perturbation theory in Sec.~\ref{sec:perturbation} and further study the
influence of an impurity on the wave packet dynamics in Sec.~\ref{sec:impurity}, followed by a brief summary of our results.

\section{Quasiperiodic Chains with Golden, Silver or Bronze Means}\label{sec:construction}

In this paper we study one-dimensional quasiperiodic systems constructed by the inflation rule
 \begin{equation}
  \label{equ:infaltion_rule}
  \mathcal{P} =
  \begin{cases}
    w \rightarrow s \\
    s \rightarrow sws^{n-1}
  \end{cases}\hspace{1cm},
 \end{equation}
iterated $a$ times starting from the symbol $w$, where the letter $w$ denotes a weak bond and $s$ a strong bond. We refer to the resulting
sequence after $a$ iterations as the $a$th order approximant $\mathcal{C}_a$ with the length $f_a$ given by the recursive equation $f_a =
f_{a-2} + n f_{a-1}$ and $f_0 = f_1 = 1$. Depending on the parameter $n$ the inflation rule generates different so-called metallic means,
i.e.\ the lengths of two successive sequences satisfy the relation
 \begin{equation*}
  \lim_{a \rightarrow \infty} \frac{f_a}{f_{a-1}} = \lambda \;,
 \end{equation*}
where $\lambda$ is an irrational number with the continued fraction representation $[\bar{n}] = [n,n,n,...]$. For example, $n=1$ yields the
well-known Fibonacci sequence related to the golden mean $\lambda_\mathrm{Au} = [\bar{1}] = (1+\sqrt{5})/2$,
the case $n=2$ corresponds to the octonacci sequence with the silver
mean $\lambda_\mathrm{Ag} = [\bar{2}] = 1+\sqrt{2}$, and for $n=3$ one obtains the bronze mean $\lambda_\mathrm{Bz} = [\bar{3}] =
(3+\sqrt{13})/2$.\cite{NonLinAnal.1999.Spinadal}

Due to the recursive inflation rule \eqref{equ:infaltion_rule}, these quasiperiodic chains possess a hierarchical structure, which is more
clearly visible by using the alternative construction rule $\mathcal{C}_a = \mathcal{C}_{a-1}\, \mathcal{C}_{a-2}\,
(\mathcal{C}_{a-1})^{n-1}$, yielding the same quasiperiodic sequences for $a \ge 2$ with $\mathcal{C}_0 = w$ and $\mathcal{C}_1 = s$. Further, for given $n$, the structure of these chains consists of
only two types of clusters with strong interactions, $s^{n}$ and $s^{n+1}$, which are separated by a single weak bond. The
latter property can be related to the eigenstates of the system, whereas the hierarchical property has a crucial influence on the transport
properties for weak coupling. We discuss both aspects in detail later.

To obtain the quantum mechanical eigenstates of these systems, we only consider nearest-neighbor hopping and hence obtain the tight-binding
Hamiltonian
 \begin{equation}
  \label{equ:hamiltonian}
  H = \sum_{l = 0}^{f_a} \ket{l} t_{l,l+1} \bra{l+1} + \sum_{l=0}^{f_a} \ket{l} \varepsilon_l \bra{l},
 \end{equation}
represented in the orthogonal basis states $\ket{l}$ associated to a vertex $l$. The off-diagonal matrix elements represent the kinetic
energy in the tight-binding model. The diagonal elements of the Hamiltonian matrix, which represent the potential energy of the sites,
are set to zero ($\varepsilon_l = 0$) because no energetic disorder is taken into account for the studied systems.

The hopping parameters $t$ are chosen according to the letters $s$ (strong) and $w$ (weak) of the quasiperiodic sequence $\mathcal{C}$ with
$t_s = 1$ and $t_w = v$ $(0 \le v \le 1)$. Thus, the number of sites of the $a$th order approximant is given by $f_a + 1$. This model can
be interpreted as describing an electron hopping from one vertex of the quasiperiodic chain to a neighboring one, and the aperiodicity is
given by the underlying quasiperiodic sequence of couplings. We then have to solve the discrete time-independent Schr\"{o}dinger equation
\begin{equation*}
 H \ket{\Psi^i} = E^i \ket{\Psi^i}
\end{equation*}
by diagonalization of the Hamiltonian matrix of Eq.~\ref{equ:hamiltonian}. Applying free boundary conditions, we obtain $f_a +
1$ eigenstates $\ket{\Psi^i} = \sum_{l=0}^{f_a} \Psi_l^i \ket{l}$ and the corresponding energy values $E^i$.

\section{Time Evolution of a Wave Packet on Quasiperiodic Chains}\label{sec:transport}

As outlined in the introduction, the dependency of transport properties on the spectral properties of the Hamiltonian is not yet fully
understood, and thus the investigation of transport properties in quasiperiodic systems continues to be of special interest. In
this section, we study the wave-packet dynamics in such systems, in particular in the limit of weak coupling where $v\ll 1$.
The results will then be related to the properties of the eigenstates using a perturbation theory approach in the
following section.

We investigate the transport properties by means of the time evolution of a wave packet $\ket{\Phi} = \sum_{l=0}^{f_a} \Phi_l \ket{l}$,
which is initially localized at the center of the quasiperiodic chain, i.e.\ $\Phi_l (t=0) = \delta_{ll_0}$ with $l_0 = \lceil f_a /2 \rceil$.
It is represented in the basis of the orthonormal eigenstates $\Phi_l (t) = \sum_i \Psi_{l_0}^i \Psi_l^i(t)$. The solutions of the
time-dependent Schrö\"{o}dinger equation then follow by the separation approach with $\Psi_l^i(t) = \Psi_l^i e^{-\mathrm{i}E^i t}$.

Besides calculating the expansion of the wave packet in space, a more detailed analysis can be obtained by the computation of the temporal
autocorrelation function (also known as return probability)
\begin{equation*}
 C(t) = \frac{1}{t} \int_0^t | \Phi_{l_0}(t') |^2\: \mathrm{d}t'
\end{equation*}
and the mean square displacement (also called the width)
\begin{equation*}
 d(t) = \left[\sum_{l=0}^{f_a} |l-l_0|^2 \, |\Phi_l(t)|^2 \right]^{\frac{1}{2}}
\end{equation*}
of the wave packet. It is known that a particle's return probability decays with a power law $C(t) \sim t^{-\delta}$, where $\delta$ is
equivalent to the scaling exponent of the local density of states,\cite{CMathPhys.1994.Holschneider, PhysRevLett.1992.Ketzmerick,
DukeMathJ.2001.Barbaroux, JStatPhys.1997.Guerin} and $\delta =1$ refers to ballistic motion. It is further known that the spreading of the
width $d(t)$ of the wave packet shows anomalous diffusion, i.e.\ $d(t)\sim t^{\beta}$ with $0<\beta <
1$.\cite{RevMathPhys.1998.SchulzBaldes, JMathPhys.1997.Roche, AdvPhys.1992.Poon, PhysRevLett.1994.Huckestein, PhysRevB.2000.Yuan} Here
$\beta=0$ corresponds to the absence of diffusion, $\beta=1/2$ to classical diffusion and $\beta=1$ to ballistic spreading.

In addition, the wave packet dynamics exhibit multiscaling, where different moments of the wave packet scale with different, nontrivially
related exponents $\beta$.\cite{EurophysLett.1993.Guarneri, JPhysA.1993.Evangelou, PhysRevB.1994.Wilkinson, PhysRevLett.1996.Piechon,
JMathPApp.2001.Barbaroux, PhysRevB.1986.Tang} While wave packet localization implies a pure point spectrum, the converse is not true, and
the more refined notion of semi-uniform localization is necessary.\cite{PhysRevLett.1995.delRio} However, the exact relations between
particle dynamics and singular or absolutely continuous spectra are less well understood. As a rule of thumb, systems with singular continuous spectra
exhibit anomalous diffusion, while absolutely continuous spectra may lead to either anomalously diffusive or ballistic
dynamics.\cite{EurophysLett.1993.Guarneri, JPhys.1995.Zhong}

\begin{figure}
 \includegraphics*[width=8.5cm]{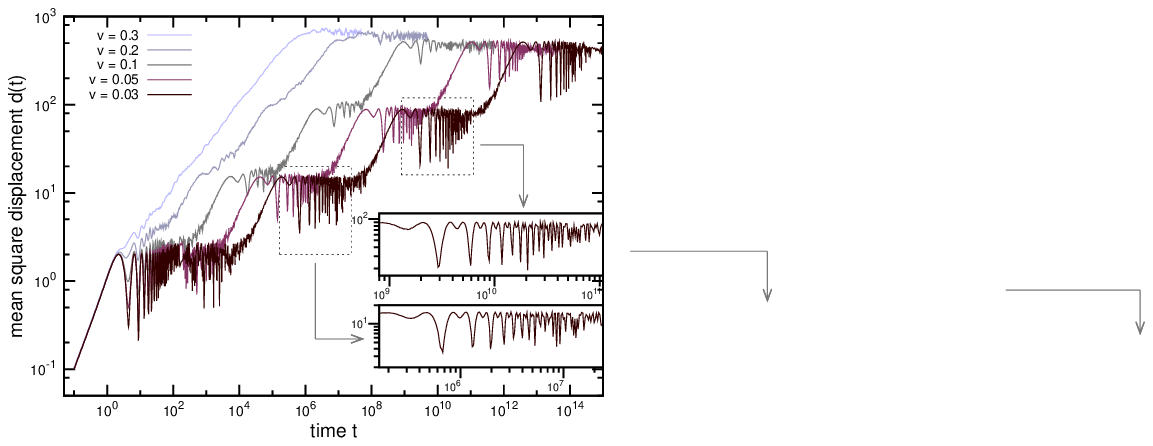}
 \caption{Evolution of the width $d(t)$ of a wave packet initially localized in the middle of the silver mean chain $\mathcal{C}_{10}^\mathrm{Ag}$ with $3364$ sites, for several small values of $v$. The insets show two magnified steps for $v=0.03$.}
 \label{fig:msd_octonacci}
\end{figure}

\begin{figure*}
 \includegraphics*[height=5.5cm]{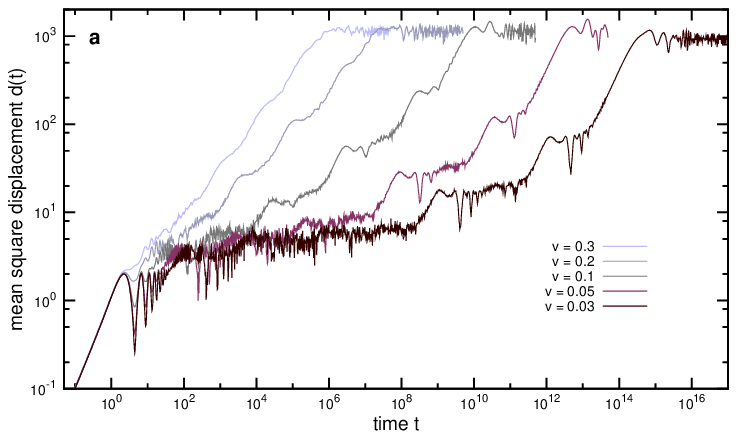}\hspace{0.3cm}
 \includegraphics*[height=5.5cm]{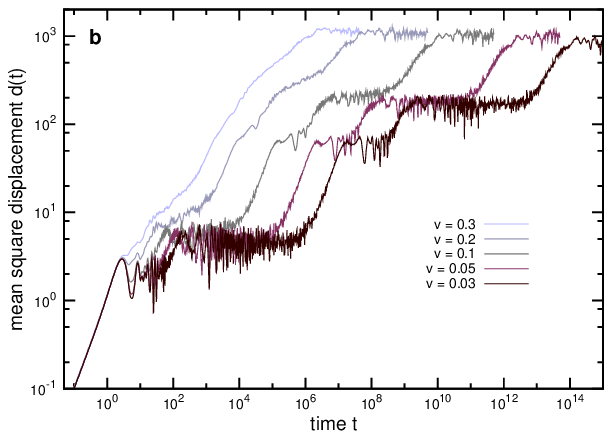}
 \caption{Same as Fig.~\ref{fig:msd_octonacci}, but for (a) the golden mean model $\mathcal{C}_{18}^{\mathrm{Au}}$ with $4182$ sites and (b) the bronze mean model $\mathcal{C}_{8}^{\mathrm{Bz}}$ with $5117$ sites.}
 \label{fig:msd_others}
\end{figure*}

\begin{figure*}
 \includegraphics*[width=5.8cm]{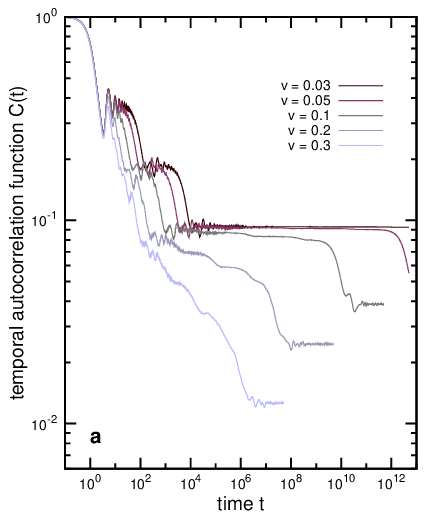}\hspace{0.1cm}
 \includegraphics*[width=5.8cm]{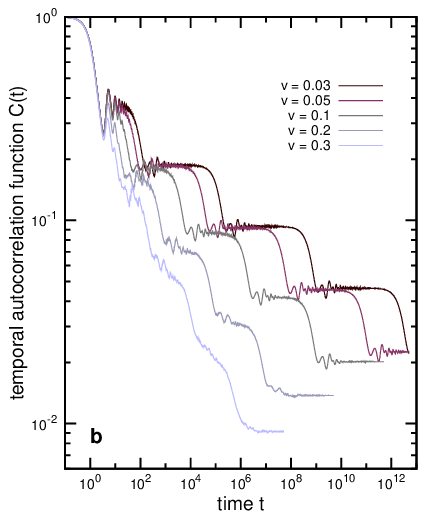}\hspace{0.1cm}
 \includegraphics*[width=5.8cm]{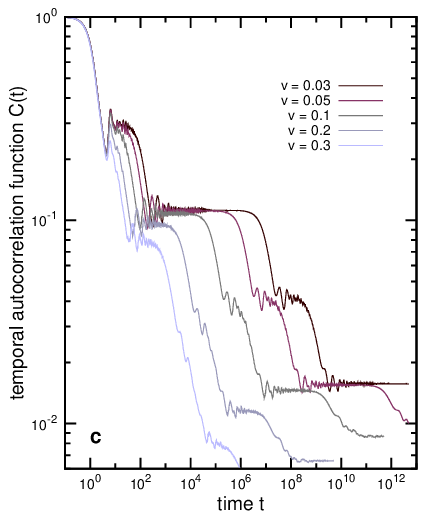}
 \caption{Temporal autocorrelation function $C(t)$ of a wave packet initially localized at the center of (a) a golden mean chain $\mathcal{C}_{18}^{\mathrm{Au}}$ with $4182$ sites, (b) a silver mean chain $\mathcal{C}_{10}^{\mathrm{Ag}}$ with $3364$ sites, and (c) a bronze mean chain $\mathcal{C}_{8}^{\mathrm{Bz}}$ with $5117$ sites. Results are shown for small coupling parameters $v$.}
 \label{fig:ac_others}
\end{figure*}

Consequently, the systems considered here should show anomalous diffusion with the mentioned power-law dependency $d(t) \sim t^\beta$ due to their singular continuous spectra,\cite{EurophysLett.1989.Sire,
PhysRevB.2000.Yuan} which is confirmed by our numerical results. Figures~\ref{fig:msd_octonacci} and \ref{fig:msd_others} show the
development of the mean square displacement of the wave packet over time for the golden, silver, and bronze mean models in the regime of
strong quasiperiodic modulation ($v \le 0.3$). While the asymptotic behavior can be characterized as anomalous diffusion, e.g.\ with scaling
exponents not larger than $\beta_{\text{Au}} \approx 0.41$, $\beta_{\text{Ag}} \approx 0.39$, and $\beta_{\text{Bz}} \approx 0.40$ obtained
for $v = 0.2$ (for higher values of $v$ and the associated values of $\beta$ see Yuan et al.\cite{PhysRevB.2000.Yuan}), there are
intervals where $d(t)$ grows according to a power law $d(t) \sim t^{\beta^\prime}$ with a constant ($v$-independent) exponent $\beta'$ which are intercepted by flat regimes.
As demonstrated by the insets in Fig.~\ref{fig:msd_octonacci}, in these flat regimes $d(t)$ strongly
oscillates in a self-similar manner, reflecting the hierarchical structure of the system. Nevertheless, the width $d(t)$ remains bounded
from above by a constant.

While for the silver mean chain the steps in $\log d(t)$ have about the same size for a particular value of $v$, we observe small as well
as large steps for the golden and bronze mean models. This may be caused by the different underlying inflation rule, which in the case of the
octonacci chain leads to a symmetric sequence (the sequence is palindromic for this case) and to asymmetric ones for the other two systems.
However, as a general trend we observe an increase of the logarithm of the time distance between successive steps with decreasing
coupling constants $v$ in all three models.

The same behavior can also be observed for the autocorrelation function $C(t)$, as shown in Fig.~\ref{fig:ac_others}. All three
quasiperiodic chains show a stepwise behavior of the return probability $C(t)$, where the minimum value of $C(t)$ depends on the system
size and on the number of $w$ bonds. Again the stepwise process is clearly visible, where a step here consists of the decrease of
$C(t)$ with a power law with $v$-independent exponent $\delta^\prime$, followed by a time interval of constant return probability. The time
intervals for the power-law behavior and the flat parts of both quantities $C(t)$ and $d(t)$ are in correspondence with each other. Further, for large
time the return probability and wave-packet width are bounded by a constant due to the finite size $f_a$ of the system.

\begin{figure}
 \includegraphics[width=8.5cm]{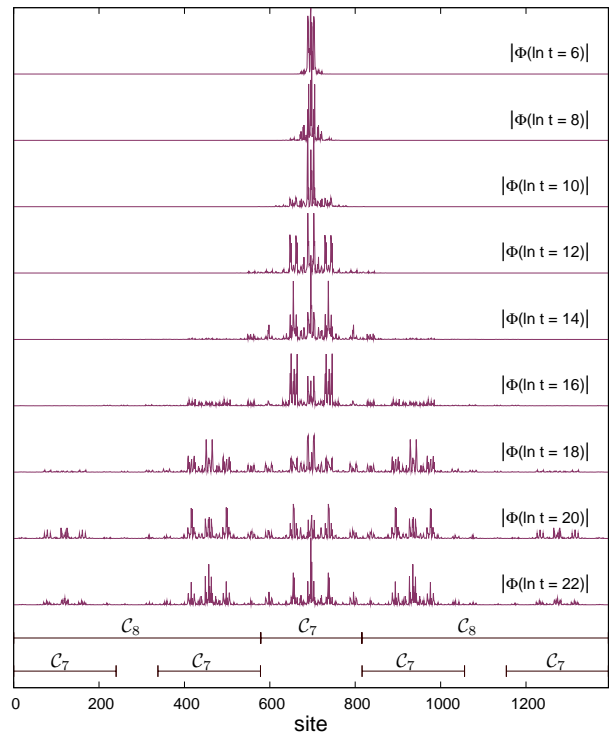}
 \caption{Snapshots for the evolution of a wave packet initially localized at the center of the octonacci chain $\mathcal{C}_9^\textrm{Ag}$ with $v = 0.1$. The wave packets are equally scaled for all considered times. Additionally, the hierarchical structure of the approximant $\mathcal{C}_9^\textrm{Ag}$ is visualized by showing the occurrence of the patterns of $\mathcal{C}_8^\textrm{Ag}$ and $\mathcal{C}_7^\textrm{Ag}$ in the chain.}
 \label{fig:wave_spreading_oct}
\end{figure}

A more detailed inspection of the wave packet dynamics reveals that breathing modes are responsible for the oscillations, while the
wave-packet spreading itself is limited to low-amplitude leaking out of the region in which it is confined. Eventually, the wave packet
expands fast to reach the next level of the hierarchy, before the whole process repeats. This behavior is even more evident in the
time evolution of the probability density of such wave packets, as shown in Fig.~\ref{fig:wave_spreading_oct}. At first up to $t = e^{14}$ in Fig.~\ref{fig:wave_spreading_oct} the wave packet is confined in a narrow range around its initial position in the environment of the approximant $\mathcal{C}_7^\textrm{Ag}$ and oscillates back and forth in this range corresponding to the strong oscillations of $d(t)$. Then, between $t = e^{16}$ and $t = e^{18}$ the wave packet expands almost ballistically and a significant probability density is found in the neighboring $\mathcal{C}_7^\textrm{Ag}$ sequences (cp.~$t \ge e^{18}$). A similar behavior can be already seen for the previous levels of the hierarchy (cp.~the panels up to $t = e^{10}$ with those for $t \ge e^{12}$ in Fig.~\ref{fig:wave_spreading_oct}), where the wave packet can be observed to spread from the central $\mathcal{C}_5^\textrm{Ag}$ structure to a three-fold $\mathcal{C}_5^\textrm{Ag}$ sequence occurring in the middle of the central $\mathcal{C}_7^\textrm{Ag}$ chain.

This indicates that the values of the flat regimes in $d(t)$ and $C(t)$ are directly related to the chain structure and do not depend on the coupling parameter $v$. In particular, we can give a rough estimate of the corresponding values of $d(t)$ for the plateaus by assuming that the wave packet is uniformly distributed in a confined region at the center of the chains. For an approximant $\mathcal{C}_a^\textrm{Ag}$ the center of the octonacci chain is made up from the three-fold sequences $\mathcal{C}_o^\textrm{Ag}\mathcal{C}_o^\textrm{Ag}\mathcal{C}_o^\textrm{Ag}$ with $o = a - 2b$ ($b \in \mathbb{N}$, $0 < o < a$). Confining the wave packet to these regions we obtain for the approximant $\mathcal{C}_{10}^\textrm{Ag}$ of Fig.~\ref{fig:msd_octonacci} the values $d(t) = 2.9, 15, 86, 500$ for $o=2,4,6,8$ respectively. Although the wave packet is far from being uniformly distributed and some parts of the wave packet can also be found outside the confined region (cp.~Fig.~\ref{fig:wave_spreading_oct}), our assumption results in a good reproduction of the plateau values observed in Fig.~\ref{fig:msd_octonacci}.

This spreading of the wave-packet width to the next level can be described by a power law $d(t) \sim t^{\beta'}$ with exponents $\beta_{\mathrm{Au}}^{\prime} \approx 0.88$, $\beta_{\mathrm{Ag}}^{\prime} \approx 0.85$, and $\beta_{\mathrm{Bz}}^{\prime} \approx 0.98$
determined for the smallest coupling constant $v = 0.03$ considered here. Computing the results also for all 6 other qualitatively different initial positions of the wave packet for the octonacci chain we obtained about the same scaling exponents $\beta^\prime$. For instance for the octonacci chain we found  $\beta^\prime = 0.81-0.85$, where some of these differences might be caused by fluctuations of the width $d(t)$ which are present in the regime of strong expansion and make fitting difficult. These scaling exponents tend towards the exponents obtained for $v\to 1$,\cite{PhysRevB.2000.Yuan} which indicates that the fast expansion is not governed by the weak coupling, but rather a kind of resonance between the different levels of the hierarchy. Further, for the return probability we find the exponents $\delta_{\mathrm{Au}}^{\prime} \approx 0.71$,
$\delta_{\mathrm{Ag}}^{\prime} \approx 0.71$, and $\delta_{\mathrm{Bz}}^{\prime} \approx 0.76$, which are again relatively close to the exponents for $v \to 1$.\cite{PhysRevB.2000.Yuan, JPhys.1995.Zhong} However, these values of the scaling exponent $\delta^{\prime}$ might differ from the exact result, because in one dimension we cannot rule out the influence of subdominant logarithmic contributions for the considered short time intervals in the step-like process.\cite{JPhys.1995.Zhong}

Similar behaviors for $d(t)$ and $C(t)$ have been reported before for the Fibonacci chain with strong quasiperiodic
oscillations.\cite{PhysRevA.1987.Abe, JPhys.1995.Zhong} Wilkinson and Austin found the same step-like process when studying the
spreading of a wave packet for Harper's equation of an electron in a magnetic field.\cite{PhysRevB.1994.Wilkinson} Based on a qualitative
model of the wave-packet spreading in the semiclassical approximation and on numerical simulations, they argued that a hierarchical
splitting of the energy spectrum into constant-width bands leads to a step-like behavior with $\beta^\prime=1$, which is smoothed due
to the (broad) distribution of band widths.

The value $\beta'<1$ observed here suggests that there is a distribution of band widths in the energy spectrum even for $v \ll 1$. The energy levels of the singular continuous spectrum for $0 < v < 1$ cluster in certain energy ranges and form bands with different widths (cp. Sec. \ref{sec:perturbation}), but all the band widths approach zero in the limit $v \to 0$.\cite{PhysRevB.1992.Passaro, PhysRevB.2005.Cerovski} This suggests that the self-similar spreading of the wave packet is only an approximate description of a more general multiscale dynamics.

Further, the self-similarity of quasiperiodic sequences was previously used in a renormalization-group perturbative expansion that provided
a great deal of insight into the eigenstate properties,\cite{PhysRevLett.1986.Niu} and showed multiscaling of wave packet
dynamics.\cite{PhysRevLett.1996.Piechon} In the following section we focus on ergodic rather than hierarchical
properties.\cite{PhysRevB.1990.Niu} By an elementary analysis of the perturbation theory of degenerate levels at $v = 0$ for small coupling
constants $v$, we show that, on the one hand, eigenstates delocalize for any $v > 0$, in contradistinction to (trivial) localization at $v =
0$. On the other hand, in the limit as $v \to 0$, eigenstates delocalize across only one set of clusters containing the same number of
atoms, i.e.\ we obtain a subcluster localization due to the breaking of ergodicity.

\begin{figure*}
 \includegraphics[width=8.5cm]{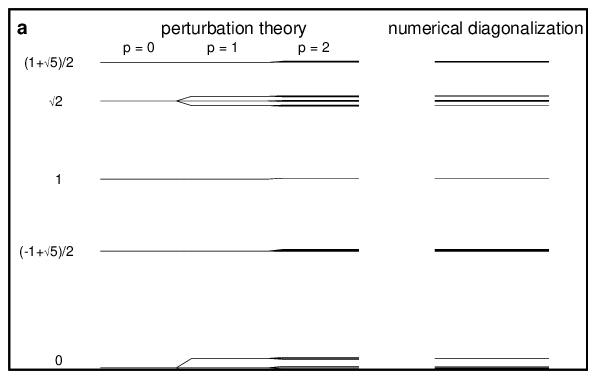}\hspace{0.5cm}
 \includegraphics[width=8.5cm]{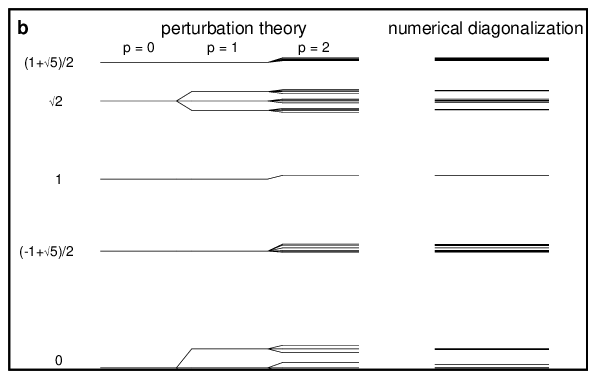}
 \caption{Comparison of energy levels obtained by the perturbation theory approach and by numerical diagonalization of the perturbed Hamiltonian $H$. Results are shown for the octonacci chain with $n = 2$, $a = 6$, and (a) $v = 0.1$ or (b) $v=0.2$. For reasons of symmetry only energy values $E > 0$ are included.}
 \label{fig:perturbation}
\end{figure*}

\section{Raleigh-Schr\"odinger theory}\label{sec:perturbation}

Raleigh-Schr\"odinger theory allows the recursive construction of matrices in subspaces for a degenerate eigenenergy to a given order
$p$, whose diagonalizations (the secular problem) yield corrections to the unperturbed eigenenergies. Within this approach, the Hamiltonian
is decomposed into an unperturbed system $H^{(0)}$ and a perturbation $H^{(1)}$ with $H = H^{(0)} + \lambda H^{(1)}$, where the hopping
parameter $v$ is treated as the perturbation yielding $H^{(0)} = H(v=0)$ and $H^{(1)} = H(v) - H (v=0)$. Although the accuracy of $\mathcal{O}(v^{p+1})$ of the expansion to the $p$th order is not guaranteed, the theory yields good results for small perturbations and
preferably large separations between the degenerate energy levels. The first condition is met since we only consider small values of $v$ and the
latter one is satisfied since we found that for the unperturbed system $|E^i -E^j| > c$ with $c^{\mathrm{Au}} = 0.41$, $c^{\mathrm{Ag}} = 0.20$ and $c^{\mathrm{Bz}} = 0.11$.

For the considered quasiperiodic sequences the results of such perturbation expansions yield two qualitatively different types of
solutions, depending on the values of $p$ and $n$ due to the chain structure mentioned in Sec.~\ref{sec:construction}. The reason is
that the chain consists of strongly coupled clusters with $n+1$ and $n+2$ atoms, which are weakly connected via the hopping parameter $v$. For
$v=0$ we have an unperturbed system with $2n+3$ highly degenerate levels, where all eigenstates are localized on individual clusters. In
higher orders of perturbation theory, these localized states then spread first over neighboring clusters of the same type as the coupling
among the clusters is taken into consideration, and, for a sufficiently high order, delocalize across the whole chain.

Since the maximal number of letters $w$ between two consecutive clusters of length $s^{n+1}$ is also $n+1$, the eigenstates of these
clusters delocalize only in order $n+1$ of the perturbation theory. However, small clusters of length $s^{n}$ are connected by at the most 2 ($n > 1$) or 3 ($n = 1$) weak bonds. More precisely, the dimension of the secular problem for each type of cluster
separated by not more than $q$ weak bonds, changes from $\mathcal{O}(q^2)$ at most for $p < q$ to $\mathcal{O}(f_a)$ for $p \ge q$. Only
the latter case allows for multifractal and/or extended states to be present in the solutions of the perturbed system.

As an illustrating example we analyze the octonacci chain with $n = 2$. In the unperturbed system there are $7$ levels, given by
$E_{sss}^{(0)}=\pm({\sqrt 5}\pm 1)/2$ and $E_{ss}^{(0)}=\pm{\sqrt 2}, 0$. Due to the sequence of the $ss$ and $sss$ clusters, the
eigenstates delocalize only in the $2$nd and $3$rd order of the perturbation theory. In first-order expansion ($p = 1$), the only
correction are $6$ levels linear in $v$, splitting off the three $E_{ss}^{(0)}$ levels, because $\,\cdot]sswss[\cdot\,$ is the only
possibility for which two clusters of the same type are connected by a single $w$ bond. There are also remaining separate $ss$ clusters, so
that the unperturbed $E_{ss}^{(0)}$ levels are still present in the spectrum. For $p = 2$, all $ss$ clusters become connected while $sss$
states are still not extended due to the existence of $\,\cdot]ssswsswsswsss[\cdot\,$ sequences in the chain. For $p = 3$, states belonging
to $sss$ levels delocalize as well. Figure \ref{fig:perturbation} shows this splitting of the energy levels, where the solutions obtained
by different orders of the perturbation theory are shown in comparison to the energy levels obtained by numerical diagonalization of the
Hamiltonian $H$. The results of the perturbation approach and the exact energy values are close, although only 2nd-order corrections are
taken into account. In the case of $E_{ss}^{(0)}$ the 2nd-order corrections seem to overcompensate the error leading to a stronger
splitting of the energy levels compared to the exact numerical results.

\begin{figure*}
 \includegraphics*[height=4.2cm]{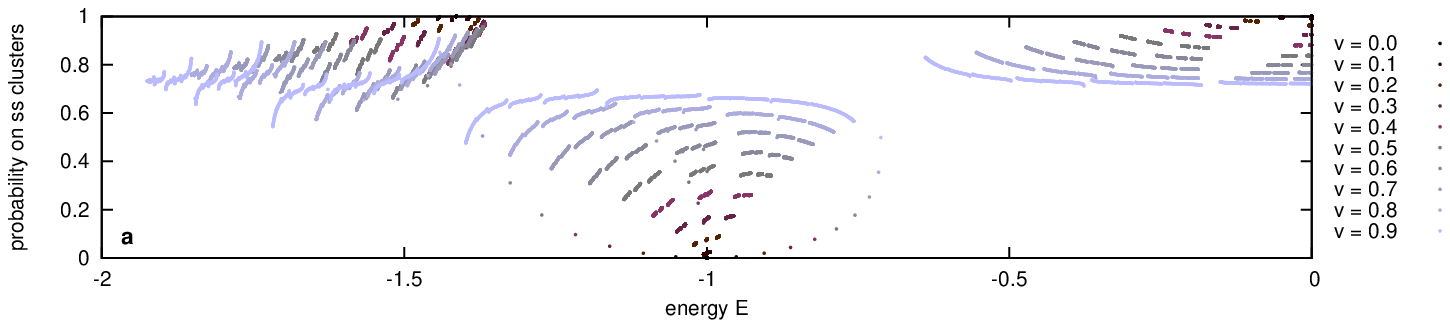}
 \includegraphics*[height=4.2cm]{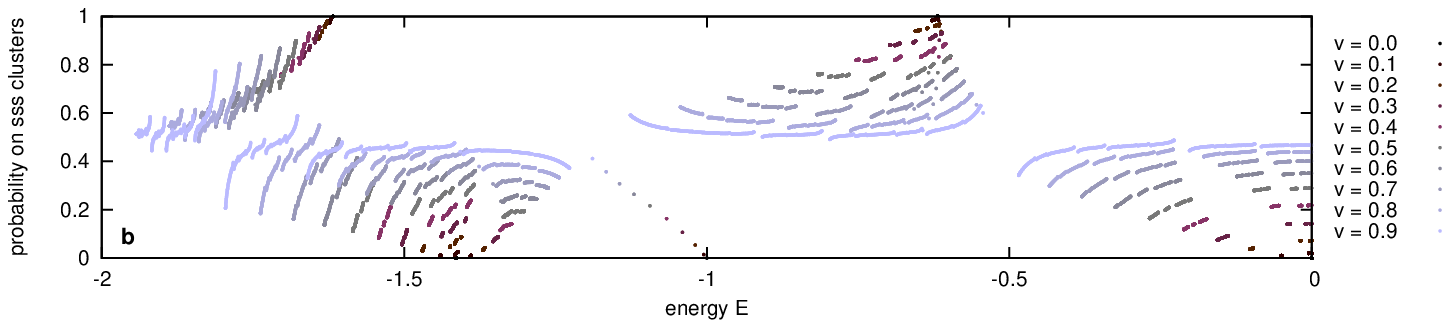}
 \includegraphics*[height=4.2cm]{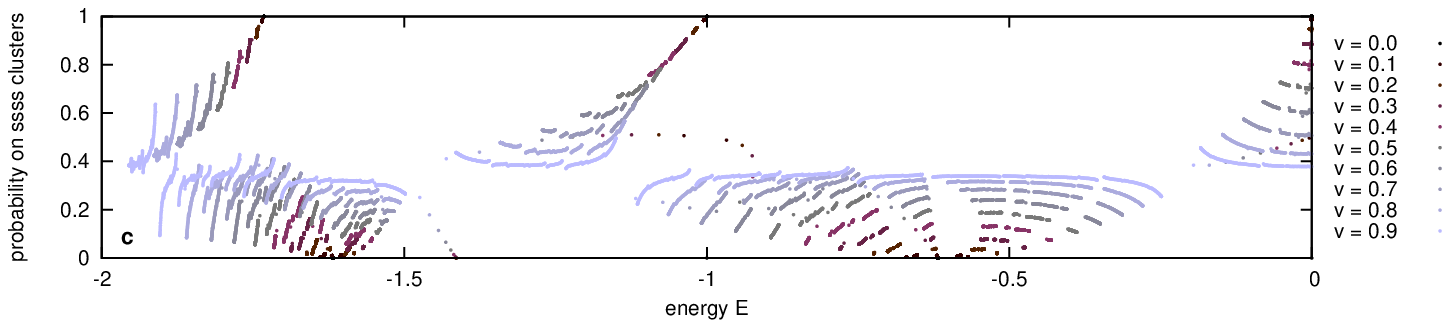}
 \caption{Probability $P_{s^{n+1}}(E)$ that a particle in an eigenstate with a given energy $E$ is on the $s^{n+1}$ sublattice, for $v=0.0,0.1,\dots ,0.9$. Only states with $E<0$ are shown due to the symmetry of the eigenstate spectrum with respect to $E=0$. Results are given for (a) the golden mean model $\mathcal{C}_{19}^{\mathrm{Au}}$ with $6765$ sites, (b) the silver mean model $\mathcal{C}_{11}^{\mathrm{Ag}}$ with $8120$ sites, and (c) the bronze mean model $\mathcal{C}_{8}^{\mathrm{Bz}}$ with $5117$ sites. Note that there are a few additional states not included in the perturbation theory approach, which are caused by the boundary conditions.}
 \label{fig:cluster_prob}
\end{figure*}

To investigate the issue of convergence further, we notice that, even when calculated to all orders of $v$, the secular problems for the
two types of clusters give solutions that are inevitably restricted to the clusters of the given type, with zero component on the
clusters of the other type. For various values of $v$ we check  by numerical diagonalization whether this can be confirmed.
Figure \ref{fig:cluster_prob} shows the total probability that the particle in an eigenstate $\Psi^i$ with an energy $E^i$ will be on a large
cluster $s^{n+1}$
 \begin{equation}
  P_{s^{n+1}}(E^i) \; = \sum_{l \in s^{n+1}} |\Psi_l^i|^2
 \end{equation}
for the golden, silver, and bronze mean model. Figure \ref{fig:cluster_prob} illustrates how the degenerate eigenvalues for $v=0$ spread into
wider and wider bands with increasing $v$. The results imply that $P_{s^{n+1}}$ strongly depends on the energy of the corresponding
eigenstate for small $v$, because it is either large for the states belonging to the $s^{n+1}$ bands and vanishes for the states of the
$s^n$ bands in the limit $v \to 0$ or shows the reverse behavior. From Fig.~\ref{fig:cluster_prob} we can also observe that there are $n+1$
bands with high probabilities and $n$ bands with low probabilities, in correspondence with the number of eigenstates generated by the large
and small clusters as obtained by our perturbation theory approach. However, for $v > 0$ the probability $P_{s^{n+1}}$ is greater than $0$
for all energy bands, which implies that the eigenstates spread over both types of clusters and thus are ergodic. Here ergodicity denotes
the spreading of eigenstates over both types of clusters.

While Fig.~\ref{fig:cluster_prob} shows results for only one iterant of the three considered models, we have also compared the results for
$P_{s^{n+1}}$ of smaller and higher approximants to see whether there is any systematic deviation from the shown values. We found that,
although there are much less states for smaller systems or many more states for larger approximants, the values of $P_{s^{n+1}}$ do not
shift systematically for almost all of the states. Instead, the additional points cluster in the same way as in Fig.~\ref{fig:cluster_prob}.

This supports the possibility that eigenstates for an infinite quasicrystalline chain remain ergodic even for very small values of $v$.
This is in disgareement with the results of the Raleigh-Schr\"odinger perturbation expansion of degenerate levels, which might,
nevertheless, still be accurate in the limit as $v\to 0$ since
 \begin{equation*}
  P_{s^{n+1}}(E^i) \xrightarrow{\,v\to 0\,}
  \begin{cases}
   1 & \text{if } E^i \text{ is caused by } s^{n+1} \text{ clusters} \\
   0 & \text{if } E^i \text{ is caused by } s^{n} \text{ clusters}
  \end{cases}\hspace{0.2cm},
 \end{equation*}
which is a necessary condition but not a sufficient one.

\section{Ergodicity and Influence of an Impurity}\label{sec:impurity}

Although the eigenstates remain ergodic even for weak coupling $v$, the wave packet dynamics are strongly influenced, as outlined in
Sec.~\ref{sec:transport}. This leads to interesting consequences when a single impurity of strength $u$ is placed at a site $l^\prime$ by changing a diagonal element of the Hamiltonian, i.e. $H^\prime = \ket{l^\prime} u \bra{l^\prime}$. At first we study the spreading of wave packets in the presence of such an impurity at different types of clusters and then present results for the maximum wave packet width when the impurity is placed at or near the initial position of the wave packet.

\begin{figure*}
 \includegraphics[width=8.5cm]{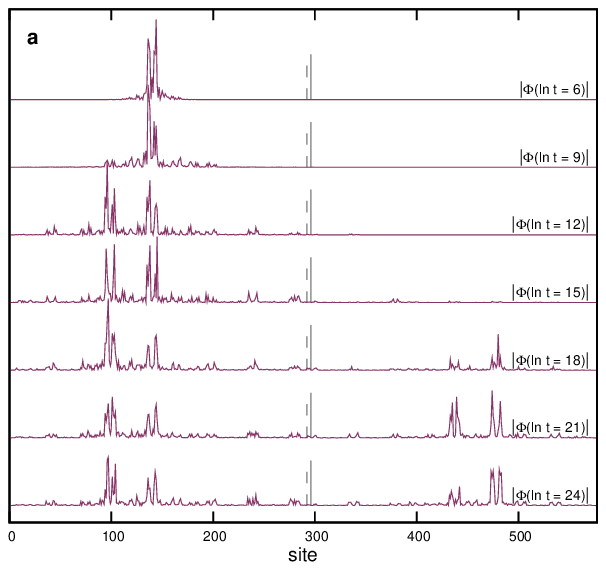}\hspace{0.5cm}
 \includegraphics[width=8.5cm]{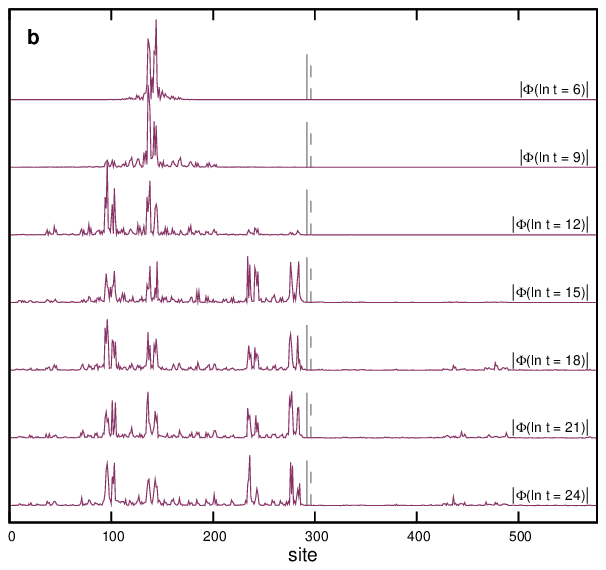}
 \caption{Snapshots of the evolution of two wave packets in the presence of an impurity for the octonacci chain with $n=2$, $a=8$, and $v = u = 0.1$. The wave packet is initially localized on a large cluster in a local environment $\,\cdot]ws_{x}ssw[\cdot\,$ at the site $x$. In the two panels (a) and (b) the impurity is located on a small cluster or a large cluster in a local environment $\,\cdot]ws_{b}ssws_{a}swsssw[\cdot\,$ at the sites indicated by $a$ and $b$, respectively, as visualized by a vertical line in each panel. For easier comparison, the vertical dashed line in each panel marks the position of the impurity in the other panel. The long-time wave packet dynamics exhibit high sensitivity on whether the impurity is located on the same type of cluster or not as the one on which the wave packet was initially localized.}
 \label{fig:wave_spreading}
\end{figure*}

In the first situation we performed several numerical experiments for various initial positions of the impurity and of the wave
packet for different values of $u$. We found that for large $u$ the impurity acts as a barrier, effectively cutting the chain into two halves. The consequence is that the wave packet is reflected at the impurity independently of its initial site, even if the coupling $v$ is small. For $u \to 0$, on the other hand, the unperturbed wave packet propagation of the case $u = 0$ is restored.

Understanding the wave packet propagation in the regime of intermediate values of $u$, however, poses significant challenges and surprising results.\cite{PhysRevLett.1995.delRio} A common situation is shown in Fig.~\ref{fig:wave_spreading} for the silver mean model, where the wave packet is initially localized at an $sss$ cluster and the impurity $u$ is placed either on an $ss$ or $sss$ cluster near the center of chain. In this case the evolution of the wave packet exhibits high sensitivity on the position of the impurity, approaching two quite different stationary states. In particular, while in Fig.~\ref{fig:wave_spreading}(a) the final state is just slightly perturbed from the final state for $u = 0$, in Fig.~\ref{fig:wave_spreading}(b) most parts of the wave packet are reflected and only a small amplitude can leak through the barrier. This kind
of wave packet dynamics is a consequence of the nearly non-ergodic spreading of the eigenstates for the two different types of clusters for
small $v$ as discussed above (cp. Fig.~\ref{fig:cluster_prob}). The explanation is that the wave packet is constructed by a superposition
of all eigenstates and because it is initially localized on an $s^{n+1}$ cluster, eigenstates caused by this type of cluster are
contributing with a much higher probability than the eigenstates of the $s^n$ clusters. Consequently, the impurity is felt as a barrier by
the wave packet when placed on the same type of cluster as the initial position of the wave packet.

\begin{figure}
 \includegraphics[width=8.5cm]{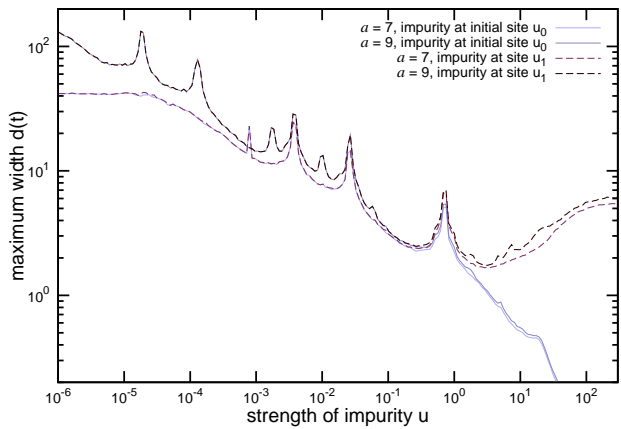}
 \caption{Maximum width of a wave packet attained during its evolution in the presence of a single impurity of strength $u$ for the octonacci chain with $n = 2$ and $v = 0.1$.  The wave packet is initially localized at the center of the chain $x = l_0$ in the local environment $\,\cdot]wssws_{u_{1\!}}s_{x}swssw[\cdot\,$ and the impurity is  either placed at the same site $u_0 = x$ or its left neighbor site $u_1$. Note that we only included approximants $\mathcal{C}_a^{\mathrm{Ag}}$ with odd $a$ because for even $a$ the local environment of the initial position of the wave packet is different.}
 \label{fig:msd_width}
\end{figure}

In the second situation we address the influence of an impurity placed at or near the initial site of the wave packet on the dynamics by studying the
dependence of the final wave packet width on the impurity strength $u$. Figure \ref{fig:msd_width} shows the maximum value of $d(t)$
attained in the course of the evolution of a wave packet in the octonacci chain, which was initially localized at the site $l_0=\lceil f_a
/ 2 \rceil$. Here only systems $\mathcal{C}_a^\mathrm{Ag}$ with odd $a$ and with the impurity placed at the initial site of the wave packet or its left neighbor site are shown. To obtain these maximum values of the width we perform the calculations as in Fig.~\ref{fig:msd_octonacci}, but now for the
perturbed system and different values of the impurity strength $u$ up to very large times, where the system is governed by finite size effects and $d(t)$ becomes constant.

The results show that, for small $u$, the width of the wave packet equals the results for the unperturbed system for both positions of the impurity. For large $u$, we obtain a strongly localized final wave packet when the impurity is added at the initial site $l_0$ of the wave packet and a constant
width of the wave packet when placing the impurity at the left neighbor site $l_0-1$. In the first case, the expansion of the wave packet in the eigenstate basis is dominated by strongly localized wave functions caused by the large impurity $u$ and the wave packet can no longer spread across the chain. In the latter case,
the impurity acts as a barrier placed at the center of the chain, and consequently the wave packet is reflected and only spreads
across one half of the system as in Fig.~\ref{fig:wave_spreading}(b), and $d(t)$ reaches a plateau in Fig. \ref{fig:msd_width}. Nevertheless, the width $d(t)$ is reduced compared to an unperturbed system with half the system size because in the presence of an impurity always some localized eigenstates, which do not spread across the quasiperiodic chain, occur and contribute to the expansion of the wave packet.

However, in between these two extremes there is a wide range of values of $u$ for which the final width of the wave packet is significantly
reduced even for
$u \ll v$, signaling dynamical localization. There are nevertheless several well-defined peaks in Fig. \ref{fig:msd_width} for some values of $u$ at which the maximum
wave-packet width is significantly enhanced, compared to the cases of slightly smaller and slightly larger values of $u$. These peaks
persist for different system sizes and for different positions of the impurity $u$, although the positions and structure of the peaks can
change. Figure \ref{fig:msd_width} shows that the peaks for the four systems considered there occur at the same strength of the impurity, especially
for relatively large values of $u$.

Having a closer look at the eigenstates caused by the impurity, we found that those peaks appear at impurities $u$, where at least some of these
perturbed states coincide with the bands in the energy spectrum of the quasiperiodic approximants shown in Fig.~\ref{fig:cluster_prob}. In this case the states caused by the
impurity hybridize with the unperturbed states of this band and consequently the wave packet is able to spread along the chain easier. This
also explains the differences of the peaks for $a=7$ and $a=9$ in Fig.~\ref{fig:msd_width}, because the structure of the energy spectrum of
the $9$th approximant is more complicated and thus additional peaks occur.

\begin{figure}
 \includegraphics[width=8.5cm]{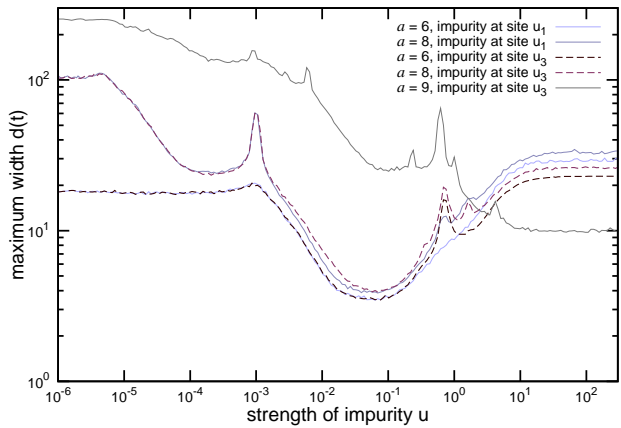}
 \caption{Same as Fig.~\ref{fig:msd_width}, but for even $a$ with the impurity of strength $u$ located at the 1st and 3rd left neighbor.
 The wave packet is initially localized at the center $x = l_0$ of the chain in the local environment $\,\cdot]wsssw_{u_{3\!}}ss_{u_{1\!}}w_{x}sswsssw[\cdot\,$.}
 \label{fig:msd_width2}
\end{figure}

Further, in Fig.~\ref{fig:msd_width2} we compare the maximum widths of the wave packet for impurities placed at the 1st left neighbor and
the 3rd left neighbor in systems $\mathcal{C}_a^{\mathrm{Ag}}$ with even $a$. It is clearly visible that the maximum widths of the wave packet for these systems are almost identical for the two different positions of the impurity. The reason is that the impurities are in both cases located at the edges of the $ss$ cluster, which is adjacent to the cluster, where the wave packet is initially localized. In contrast the curve of the approximant
$\mathcal{C}_9^\mathrm{Ag}$ shows a completely different behavior, because here the wave packet is located in a different local environment
in the beginning.

Further, by repeating similar numerical experiments for various values of the coupling strength $v$, we found that the peak structure becomes less distinct with increasing $v$. For the silver mean model we observed that it persists up to $v \approx 0.4$. For this value of $v$ the
widths of the energy bands become smaller than the gaps between them, which in turn means that the impurity-related eigenstates coincide very often with the energy bands and thus peaks merge so that almost no valleys occur.

These results show that in quasiperiodic quantum wires one can strongly influence the long-range electronic transport properties by
inducing local perturbations at different positions and of various strengths. The characteristics are related to the nature of the eigenstates, which spread only across one type of cluster in the limit $v \to 0$. Knowing the structure of the energy bands and of the eigenstates allows one to design quasiperiodic chains with impurities that can act as sort of control gates.

\section{Conclusion}\label{sec:conclusion}

In this paper, we considered the electronic transport in one-dimensional quasiperiodic chains consisting of two types of clusters
which are weakly
coupled by a hopping parameter $v$. The investigations of the wave-packet dynamics revealed the occurrence of a stepwise process with time
intervals of power-law growth, followed by a regime with confined wave-packet width. Nevertheless, the average wave packet dynamics can be classified as
anomalous diffusion. These results are consistent with the literature.\cite{PhysRevB.2000.Yuan, JPhys.1995.Zhong, PhysRevB.1994.Wilkinson}
The stepwise behavior is caused by the hierarchical structure of the chains, leading to the confinement of the wave packet until it expands
fast to reach the next level of the hierarchy.

The perturbation theory approach allowed us to draw a connection between the structure of the weakly coupled clusters and the
localization characteristics of the eigenfunctions $\Psi$, which only become delocalized in sufficiently high orders of the perturbation
expansion. This happens when clusters of a specific type become connected, i.e.\ in order $n+1$ of the expansion for large clusters and in
the 2nd or 3rd order for small clusters.

However, while we obtained that the eigenfunctions spread ergodically over all clusters of the chain for $v > 0$, in the limit $v \to 0$ we
found that ergodicity is broken and the eigenfunctions can only spread across clusters of the same type. The latter case has a significant
influence on the long-term wave packet dynamics in the presence of local perturbations. We showed that the initial site determines whether
the wave packet is reflected at the impurity or whether it can tunnel through the impurity, and that the maximum width of the wave packet
is affected by the strength and the position of the impurity. This behavior might be used to construct control gates for the electronic transport in quasiperiodically modulated quantum wires.

Further research is planned in the following directions: It is intended to consider the properties of more complex systems, as e.g.~the associated labyrinth tilings in two and three dimensions. Further, instead of investigating an on-site disorder it would be also interesting to study the influence of phason disorder caused by exchanging certain strong and weak bonds.

\begin{acknowledgments}
The authors thank V.\ Cerovski for his preliminary work on this topic and the Stiftung der Deutschen Wirtschaft and EPSRC (grant EP/D058465)
for funding the research.
\end{acknowledgments}

\newcommand{\noopsort}[1]{} \newcommand{\printfirst}[2]{#1}
  \newcommand{\singleletter}[1]{#1} \newcommand{\switchargs}[2]{#2#1}

\end{document}